# Exploring Direct Citations between Citing Publications

Yong Huang

*Information Retrieval and Knowledge Mining Laboratory, School of Information Management, Wuhan University, Wuhan, Hubei, China*

Yi Bu

*Department of Information Management, Peking University, Beijing, China*
*Center for Complex Networks and Systems Research, Luddy School of Informatics, Computing, and Engineering, Indiana University, Bloomington, IN., U.S.A.*

Ying Ding

*School of Information, University of Texas, Austin, TX, U.S.A.*
*Dell Medical School, University of Texas, Austin, TX, U.S.A.*

Wei Lu

*Information Retrieval and Knowledge Mining Laboratory, School of Information Management, Wuhan University, Wuhan, Hubei, China*

**Notes:**

1. **Yong Huang and Yi Bu contributed to this article equally.**
2. **Corresponding author: Wei Lu (weilu@whu.edu.cn).**

**Abstract**: This paper defines and explores the direct citations between citing publications (DCCPs) of a publication. We construct an ego-centered citation network for each paper that contains all of its citing papers and itself, as well as the citation relationships among them. By utilizing a large-scale scholarly dataset from the computer science field in the Microsoft Academic Graph (MAG-CS) dataset, we find that DCCPs exist universally in medium and highly cited papers. For those papers that have DCCPs, DCCPs do occur frequently; highly cited papers tend to contain more DCCPs than others. Meanwhile, the number of DCCPs of papers published in different years does not vary dramatically. The current paper also discusses the relationship between DCCPs and some indirect citation relationships (e.g., co-citation and bibliographic coupling).

## INTRODUCTION

Citation counts play a dominant role in evaluating the impact of scientific papers, researchers, related venues (e.g., journals, academic conferences), institutions, and countries [1-4]. The assessments of research grants, peer judgments, and academic ranks are often closely correlated to citation counts [5-8]. Although this indicator is simple to calculate, criticism of using citation counts in scientific assessments has also appeared in the past decades [9-12], one of which is that the assumption that every citation is equal seems problematic. To this end, bibliometricians have invested great effort into distinguishing different citations, such as assigning different weights to self-citations [13] and utilizing PageRank as an alternative to citations [14,15].

Although differentiating citations and considering them separately, these improvements still utilized a single *number* to represent citations without considering details indicated by this number. Indeed, citation numbers themselves could not capture certain essential information behind this single number, such as the temporal pattern upon which these citations have been accumulated [16] and how impact differs among citing papers [17].

Among these problems that a limited number of previous studies have addressed is the nuances of citation relationships among citing papers of a given paper. Clough, Gollings, Loach, and Evans [18] built up a new citation network by keeping the longest path between two nodes (papers) and removing all other edges (citation relationships); this procedure is called "transitive reduction," aiming to remove all of the "unnecessary" edges for the flow of information to be maintained (p. 190). The newly designed network contains a central paper (also called a focal paper), all citing papers, and filtered citation relationships between the central paper and its citing papers, as well as those between citing papers. Nevertheless, the study of Clough *et al.* [18] purely compared network-level differences between the original and the newly constructed networks without in-depth discussions and explorations.

Hence, different from the raw number of citations that simply counts the number of times a paper has been cited by other papers (as shown in Figure 1(a)), in the current paper, we explore direct citations between citing publications (DCCPs) by defining an ego-centered citation network for each paper (shown in Figure 1(b)) and by considering the citation relationships: (1) between the paper and its citing papers; and (2) between a paper's citing papers. In an ego-centered citation network, such as that presented in Figure 1, each node represents a paper, and a directed edge from the source node to the target node serves as the citation relationship from the citing to cited papers. Two types of edges are distinguished in Figure 1(b): a solid line shows the direct citation relationships (DCRs) of paper *A* (the "central" [focal] paper), while a dotted line indicates the direct citations between citing publications (DCCPs), i.e., citation relationships between paper *A*'s citing papers.

The ego-centered citation networks not only quantify the number of citations (one single number) but also examine the whole structure of its citation networks (a whole directed network containing rich information, e.g., network centrality, clustering, in- and out-degree, and network dynamics when a temporal view is considered). Thus, the proposed ego-centered citation networks might be employed as a tool for research assessment and general citation analyses.

Meanwhile, the ego-centered network proposed in the current work implicitly contains both direct and indirect citations (e.g., co-citation and bibliographic couplings). Thus, one of the potential applications of the ego-centered citation networks is to quantify the semantic similarity between entities (e.g., papers, authors, journals, affiliations, etc.) by calculating the weighted distance derived from the hybrid scholarly networks. Practically, the network could combine multiple scholarly relationships, such as citation, co-citation, bibliographic coupling, co-authorship, and even co-word [19]. These show a few of the many potential applications of the ego-centered citation network, which can be deeply understood by DCCPs.

Therefore, the current paper defines DCCPs and addresses the following two related research questions: (1) What is the frequency of the occurrence of DCCPs?; (2) How do the number of DCCPs differ for papers with different citation impacts and in different years? This paper answers these two research questions particularly in the field of computer science. Thus, we should note that the questions are not answered universally as answering these research questions require using a set of scientific publications from multiple disciplines in the empirical study.

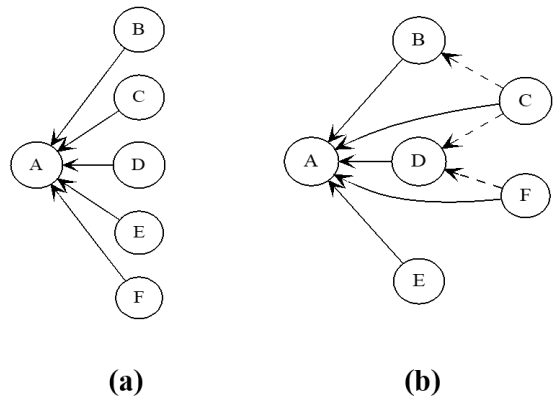

(a)                    (b)

**Figure 1.** Illustrations of the raw number of citations (sub-figure a) and our proposed ego-centered citation network (subfigure b) considering direct citations between citing papers (DCCPs). Note that both solid and dotted lines are direct citations, but the dotted lines highlight those between the citing papers of *A*, while solid lines emphasize those between *A* and its citing papers.

In the following section, how our proposed ego-centered citation network relates to some commonly discussed scholarly relationships (e.g., co-citation and bibliographic coupling) is illustrated. The dataset and the method are introduced in the methodology section. A detailed explanation about the above two research questions is presented in the result section. Discussions, implications, limitations, and future work are presented in the final section.

## DCCP AND SCHOLARLY RELATIONSHIPS

In our proposed ego-centered citation network presented in Figure 1(b), there are many direct and indirect citation relationships. Scientometrically, a direct citation from paper B to paper A implies that A occurs in the reference list of B. There are at least three branches of research focusing on direct citations from the perspectives of bibliometric indicators [3,15], citing behavior [20], and knowledge flow [21]. Direct citations have been widely used in various fields, such as scientific evaluation, information retrieval, and knowledge (innovation) diffusion (e.g., [22,23]).

Indirect citations between papers A and B refer to a non-citation relationship between A and B but imply citation relationships between A and another paper (say C), as well as between B and C. Research on indirect citations started in the 1960s, when Kessler [24] first proposed the term "bibliographic coupling" (BC) to represent the fact that two papers cite common reference(s), and found that the more shared references two papers possess—defined as greater "bibliographic coupling strength"—, the greater possibilities that they have more topical relatedness. Starting from then, BC has become an important scholarly relationship in scientometrics [25,26]. Several decades later, Zhao and Strotmann [27,28] applied BC to author levels and proposed author bibliographic coupling analysis (ABCA); they argued that ABCA tends to show more research frontiers in knowledge domain mappings.

Symmetrically, if two papers are cited by common papers, their relationships are named "co-citation" [29]. Co-citation analysis (CA) assumes that the more two papers are co-cited, the more topical relatedness they tend to have. CA has been expanded on many bibliometric entities, such as authors (author co-citation analysis [ACA], e.g., [30,31]) and journals (journal co-citation analysis [JCA], e.g., [32]) to better depict scientific intellectual structures and map knowledge domains. Studies using CA to map knowledge domains are much more than those with BC, partly because CA depicts a dynamic picture while BC tends to be a static one on paper level analyses—the co-citation frequency of two publications might change over the years, but the bibliographic coupling strength does not when a publication-level analysis is implemented. Recent efforts on co-citation analysis include adding citing contents to co-citation analysis to detect the nuance of knowledge mapping. For instance, Jeong, Song, and Ding [33] involved content information about the co-occurred citations to show an improved ACA mapping in a domain which traditional ACA failed to identify; specifically, they defined the cosine similarity between citing sentences containing two authors' publications as their co-citation frequency. More recently, Yu [34] proposed author tri-citations, defined as three authors cited by the same publication; her proposed strategies are found to improve the quality of knowledge domain mappings.

We have introduced many details on indirect citations. But how do DCCPs relate to indirect citations? In an ego-centered citation network shown in Figure 1(b), besides direct citations, the relation between papers $A$ and $B$ is also a co-citation relationship from the perspective of $C$ [29-32], given that paper $C$ cited both $A$ and $B$. However, most co-citation analysis studies (e.g., [35-37]) do not distinguish potential differences between edges $C \to A$ and $C \to B$. For example, in the current study, $C \to B$ is defined as a DCCP from the perspective of $A$, while $C \to A$ is a direct citation relationship. This indicates that co-citation links should be treated differently in various scenarios. Essentially the DCCPs reveal the *asymmetries* of co-citation relationships. Moreover, indirect citations focus more on coupled and co-cited literature (and other bibliometric entities; the same below) instead of the literature that triggers such co-citation.

The citation network shown in Figure 1(b) is ego-centered. As introduced and discussed by White [38], ego-centered citation analysis could be utilized as a useful tool to understand an author's coauthors, citation identity (i.e., his/her cited authors), citation image makers (i.e., his/her citing authors), and citation image (i.e., authors co-cited with him/her) [39]. There are two differences between White's ego-centered network [38] and ours. On the one hand, White considered both co-authorships and citation relationships, while ours considers the latter. We also focus on publication-level, but he focused on author-level, relationships. On the other hand, White did not include any citation relationships between citing authors (i.e., indirect citation relationships), but we take into consideration citation relationships between citing publications. These differences should attribute to distinct research objectives between White's and ours. Thus, the current proposal constitutes an important extension of White's framework by considering more "interactions" (citation relationships) between citing papers[1].

---

[1] If replacing citation relationships in the network with co-authorships, one can depict a more nuanced picture of the collaboration patterns among an author's co-authors. For instance, $C \to B$, $C \to A$, and $B \to A$ show the transitivity of co-authorships. If we consider more attributes of co-authors (e.g., whether they come from the same affiliation, whether they have received a similar number of citations, and whether they are in the same gender, etc.), one can also measure their homophily [40].

## METHODOLOGY

*Data*

The dataset used in the current study is the Microsoft Academic Graph (MAG) [41], which has been used and evaluated in many previous studies [42,43]. We selected all publication records in the field of computer science published from 1970 to 2016, notated as MAG-CS. The reasons why we select computer science in our empirical study lie in threefold. First of all, computer science is a rapidly developing field in the current era with many of its sub-field emerging (e.g., artificial intelligence, robotics, and security), and, therefore, understanding citation patterns of publications in computer science assists us to better learn computer scientists' citing behavior. Secondly, computer science is a field that features a large volume of papers these years, which offers bibliometricians excellent scholarly datasets to quantitatively study citation patterns in this field. Thirdly, topically computer science is adjacent with information science (i.e., the field of authors of this paper); such proximity makes it easier for us to interpret empirical results, if needed.

Figure 2 shows how the number of publications changes over time, and it can be seen that the number of papers increases steadily over years, except for the year of 2016 (data incompletion issue). The total number of papers in the MAG-CS dataset is 7,899,617. Out of these, there are 3,192,615 computer science papers that were cited at least once. By including these 3M+ papers' citing papers in the MAG dataset, and citing relations between these citing papers, a citation network comprising 40,790,926 citation relationships was built. Obviously, some of the citation relationships are not only from CS but also other fields in MAG—for instance, a CS paper might be cited by two papers from outside domains.

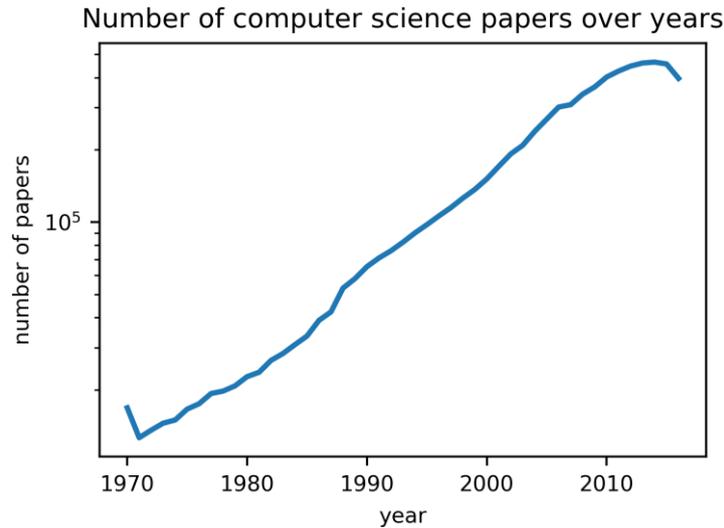

**Figure 2.** Distribution of the number of publications over time. Data source: Microsoft Academic Graph (MAG).

*Ego-centered citation networks*

For each paper, we construct an ego-centered citation network using Python automatic scripts. In addition to considering the citation relationships between a paper and its citing papers (such as that shown in Figure 1(a)), we also include citation relationships among its citing papers, as shown in Figure 1(b). Obviously, our defined ego-centered citation network of a given paper contains two types of citation relationships (edges). The first type of citation relationships shows the direct connection between a paper with one of its citing papers (e.g., the edge from nodes $B$ and $A$), illustrated as solid lines in Figure 1(b); these are direct citation relationships (DCRs). The other type of relation, presented as dotted lines, links two of the citing papers (e.g., the edge from nodes $C$ to $B$) if one paper cites another, i.e., DCCPs. Meanwhile, in this paper, we term the focal papers (node $A$ in Figures 1(a) and 1(b)) as the *owner* of the ego-centered citation network.

Due to the availability of pre-print and early view of publications, we detect some loops in certain ego-centered networks (e.g., some *arXiv* papers might be cited by paper A, and this *arXiv* paper also cites A in its later revision [44,45]; from a retrospective aspect, we will see a loop between A and this *arXiv* paper in record). In practice, we simply remove the whole ego-centered citation network if it contains one or more loops. After this process, we have a total of 2,855,035 ego-centered citation networks in our dataset.

*Measurements*

To quantify the frequency of DCCP occurrences, we calculate the probability of an ego-centered citation network whose edge count was larger than the citation count of the owner. *An ego-centered network with an edge count larger than its citation count must have at least one DCCP*. Let $C$ refer to the number of citations of a paper. Mathematically, in a specific ego-centered citation network with $(n + 1)$ nodes (i.e., the owner has been cited $n$ times) and $e$ edges ($e_d$ DCRs and $e_i$ DCCPs, and thus $e_d + e_i = e$), this probability, annotated as $P(e > n \mid C = n)$, is defined as follows:

$$P(e > n \mid C = n) = \frac{N(e > n \mid C = n)}{N(C = n)} \tag{1}$$

where $N(e > n \mid C = n)$ is the number of ego-centered citation networks with an edge count larger than the citation count, given a specific number of citations received; and $N(C = n)$ is the number of ego-centered citation networks whose owners received $n$ citations. For instance, in our dataset, we find that there were 56,482 papers that received 10 citations so far, among which there were 46,294 papers whose eco-centered citation networks had more edges than the citation count of the paper. Hence, $P(e > n \mid C = 10)$ equals $\frac{46,294}{56,482} \approx 81.96\%$.

Mathematically, it is simple to determine that $P(e > n \mid C = n)$ also equals $P(e_i > 0 \mid C = n)$ and $P(e_d \neq e \mid C = n)$. A higher $P(e > n \mid C = n)$ value indicates a high ratio of DCCPs existing in the ego-centered networks, given a specific number of citations received ($C = n$).

We also calculate the relative number of DCCPs as a robustness test, i.e., the ratio between the DCCP count in the ego-centered citation network and the number of citations of the paper (DCR count). Mathematically, in an ego-centered citation network with $(n + 1)$ nodes (i.e., the owner has been cited $n$ times), let $e_i$ be the number of DCCPs and $e_d$ be the number of DCRs. The relative number of DCCPs, $e_{i-norm}$, is calculated as:

$$e_{i-norm} = \frac{e_i}{e_d} \qquad (2)$$

Note that $e_{i-norm}$ is also equivalent to $\frac{e}{n} - 1$ ($= \frac{e-e_d}{e_d}$ where $e_d = n$), given $e$ as the total number of edges in the network ($e = e_i + e_d$) and $n$ as the paper's citation count. This indicator is straightforward. An ego-centered citation network with greater $e_{i-norm}$ reveals that it has relatively more DCCPs. We know that a network without any DCCPs will have $e = n$, and thus $e_{i-norm} = 0$. $e_{i-norm} = 1$ reveals that in an ego-centered citation network, the number of DCCPs is identical to the number of DCRs (i.e., the number of citations of this owner).

*Paper partitioning strategy*

To answer our research questions, we divide all papers into three groups based upon their numbers of citations, namely highly, medium, and lowly cited papers and observe how the number of DCCPs was distributed among these three paper types. We here adopt the strategy proposed by Guo, Milojević, and Liu [46] that partition papers with different citation counts according to their citation distributions and define highly, medium, and lowly cited papers as those whose citation counts were $[96, +\infty)$, $[23,96)$, and $[0,23)$ for our dataset, respectively.

# RESULTS

Figure 3 presents the relationship between papers with different numbers of citations and their $P(e > n \mid C = n)$ based on the MAG-CS dataset. Initially, we found that the value of $P(e > n \mid C = n)$ was low when the citation count was small, revealing that there were many lowly cited papers having no DCCPs in their ego-centered citation networks. Nevertheless, $P(e > n \mid C = n)$ increased rapidly as papers' citation count increased. When the citation count reached 10, 81.96% of the papers in the MAG-CS dataset had DCCPs in their ego-centered networks. For papers with citation counts greater than 30, $P(e > n \mid C = n)$ is close to 1.0, indicating that almost all of the papers with 30 or more citations in the computer science field featured DCCPs in their ego-centered citation networks. Hence, we can conclude that although papers with a limited number of citations do not tend to have any DCCPs, those with great numbers of citations are more likely to have DCCPs than those that have fewer citations. Moreover, since medium and highly cited papers must have had 10+ citations (actually they have 23 or more based on the aforementioned strategy [46]), we know that DCCPs exist universally in medium and highly cited papers, with $P(e > n \mid C = n)$ rapidly approaching 1.0 (see Figure 3).

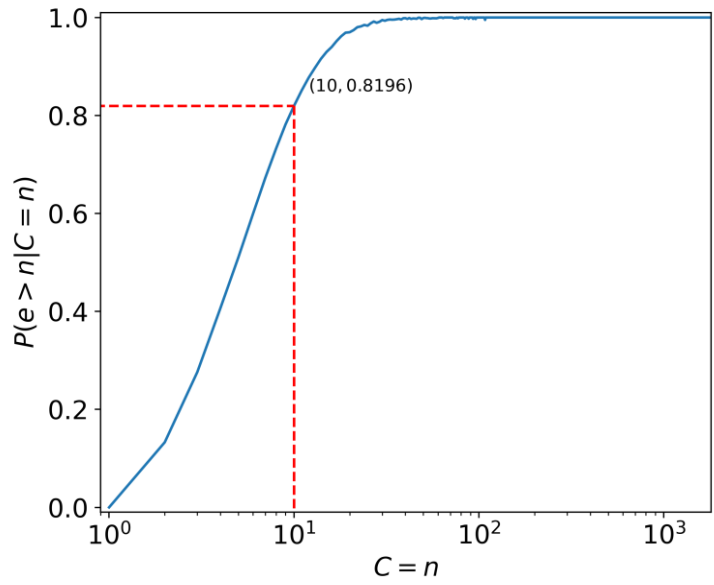

**Figure 3.** Trends of $P(e > n \mid C = n)$ with the increase of citation count.

Figure 4 shows the complementary cumulative distribution function (CCDF) of $e_{i-norm}$ among all papers with at least one DCCP, regardless of their numbers of citations, in the MAG-CS dataset. One can see that the curve of CCDF exhibits a decreasing trend when the value of $e_{i-norm}$ increases. We can find that approximately 20% of the papers with DCCPs have $e_{i-norm} > 1$, revealing that these papers have more DCCPs in their ego-centered citation networks than their own number of citations. In addition, approximately 80% of the papers with DCCPs have a number of DCCPs that is exactly or more than one-fifth of their citation count. This finding shows that DCCPs occur frequently in papers' ego-centered citation networks.

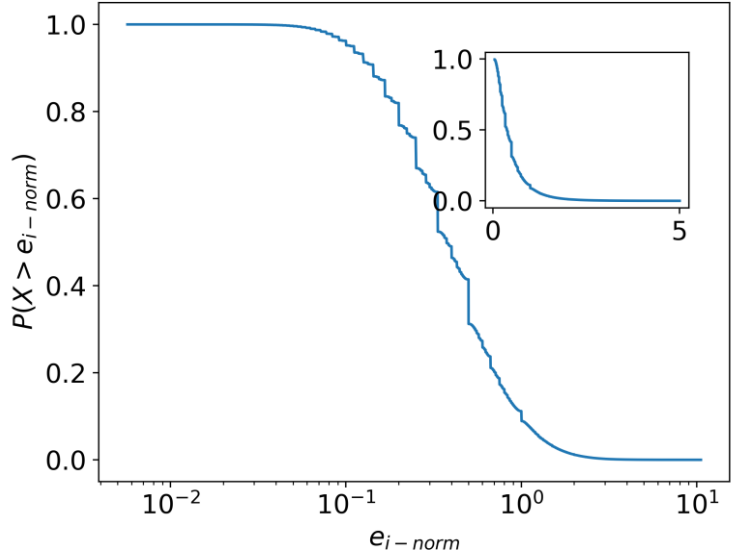

**Figure 4.** Complementary cumulative distribution function (CCDF) of $e_{i-norm}$ (all papers without any DCCPs have been removed). The small figure in the upper-right corner presents the same CCDF but the horizontal axis is normal instead of logarithmic.

We calculate the maximum, mean, and minimum values of $e_{i-norm}$ for lowly, medium, and highly cited papers, as shown in Figure 5. The top and bottom values correspond to the maximum and minimum values of the group, respectively, while the orange lines in the middle refer to the mean values. From Figure 5, a clear increasing trend can be seen of the $e_{i-norm}$ mean value as papers' citation counts increase. Specifically, the mean of the highly cited paper group is ~1.21, indicating that for these papers, on average, the number of DCCPs is 21% more than the number of citations. For medium and lowly cited papers, the means of $e_{i-norm}$ are approximately 0.77 and 0.44, respectively, which means that the numbers of DCCPs in their ego-centered citation networks are approximately 77% and 44% of their numbers of citations, respectively.

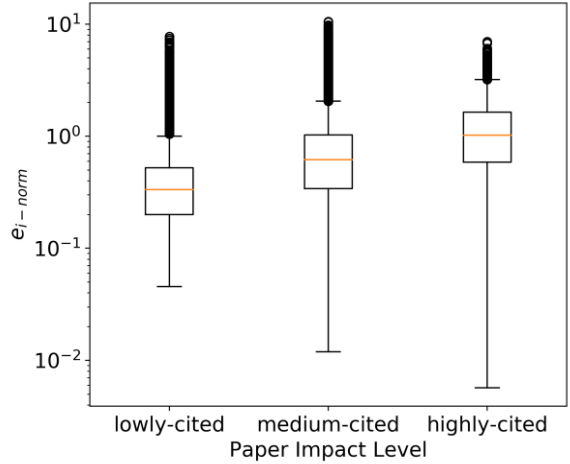

**Figure 5.** Maximum, mean, and minimum values of $e_{i-norm}$ for lowly, medium, and highly cited paper groups. All papers whose number of DCCPs equals zero have been removed.

These findings show that DCCPs frequently occur in papers' ego-centered citation networks, with highly cited papers, corresponding to higher citation impact, exhibiting the most. One of the interpretations of this finding is attributable to researchers' literature retrieval behavior. Previous empirical studies have found that "snowballing" constitutes an effective approach to find related literature in research [47,48]. This approach assists researchers to identify relevant publications by searching reference lists of previously retrieved studies [30,49]. As the current dataset prevents us from studying this process more in-depth, this interpretation implies how the number of citations accumulates with the increasing of DCCPs.

We then create a scatter plot between the relative number of DCCPs and citation counts for the papers, as shown in Figure 6. Note that the color of a dot represents the number of papers with the corresponding relative number of DCCPs and citation counts. The color bars are shown in the right of the figure. One can see that the dots in the lower-left part tend to be orange, and those in the upper-right part tend to be blue. Moreover, we find that the number of papers with a lower citation count and relatively fewer

DCCPs is much more than that with a higher citation count and relatively more DCCPs.

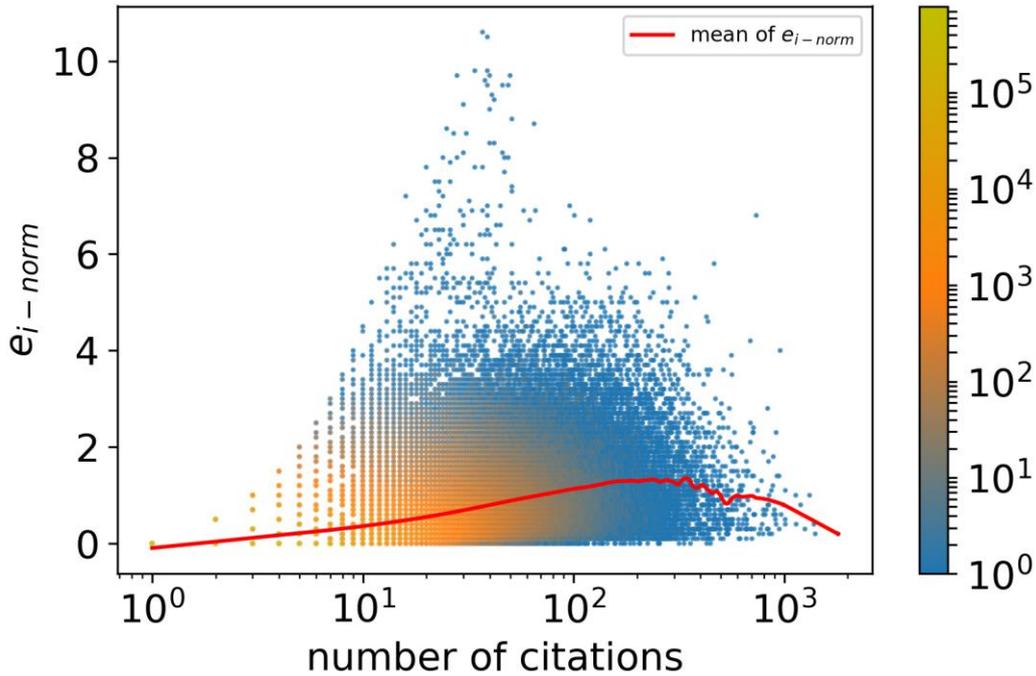

**Figure 6.** Relationship between the relative number of DCCPs ($e_{i-norm}$) and citation counts. The red curve shows the mean curve for the relative number of DCCPs.

To elucidate the relationships between the number of DCCPs and citation count, we also plot the mean value of relative number of DCCPs with the number of citations in a red curve. We find a roughly increasing trend, again, indicating that the number of DCCPs of highly and medium cited publications is greater than lowly cited publications.

In our dataset, the published years of papers range from 1970 to 2016. An intuitive question is whether the number of DCCPs for papers published in different years changes over decades. This also serves as the robustness test for our aforementioned empirical results. Similar to Figure 6, we here separately analyze high-, medium-, and low-impact papers (i.e., highly, medium, and lowly cited papers) and present both heat scatter plots and average value lines in Figure 7. From the perspective of heat scatter plots, it can clearly be seen in the sub-figures indicating lowly and medium cited paper groups that the lower-right part features some orange points, which is consistent with the results shown in Figure 2 that there are more papers published in recent years. Regarding lines indicating the average value of groups, in general one can see that the relative number of DCCPs does not change obviously for lowly and medium cited papers published in different years. The average values of the relative number of DCCPs for the two groups are 0-1 and ~1, respectively. Also, highly cited papers published before 2010 show stable $e_{i-norm}$ values, indicating the stability of researchers' citation behavior over time, despite that researchers tended to include more references in a single paper [41]. The fluctuation for publications after 2010 is partly because of the coverage of our dataset. The average values of the relative number of DCCPs of highly cited papers are between one and two, which is higher than the other two groups. This result shows the robustness and is consistent with that shown in Figure 5.

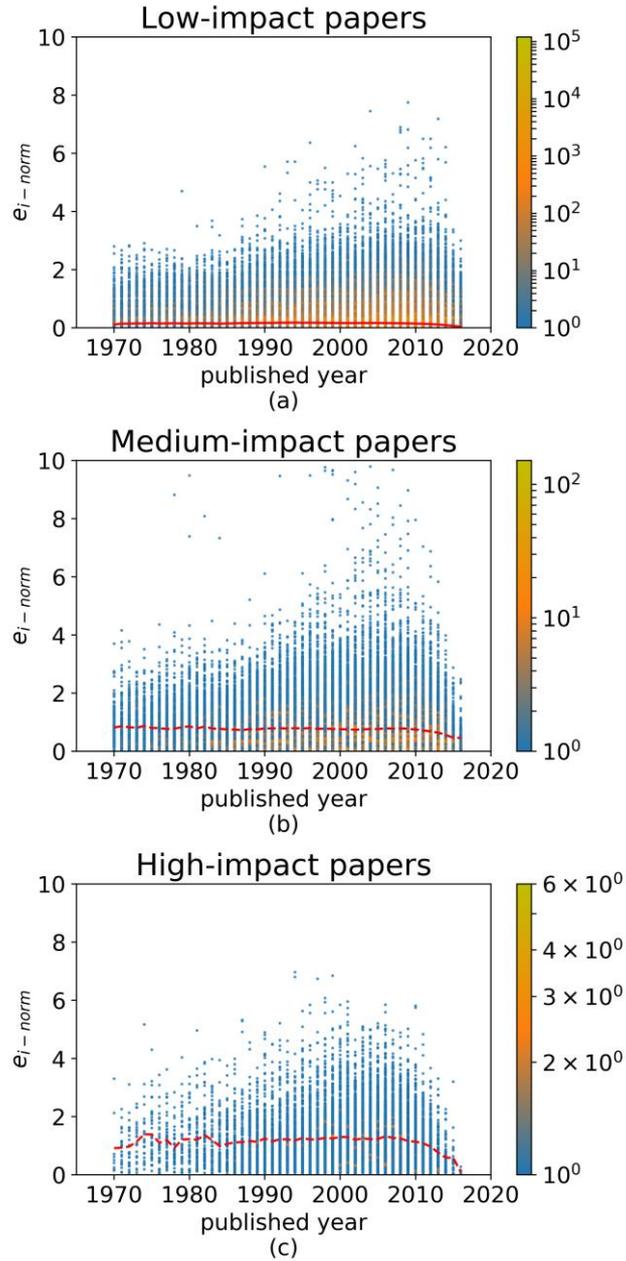

**Figure 7.** How the relative number of DCCPs ($e_{i-norm}$) of papers published in different years change over decades for (a) lowly cited papers, (b) medium cited papers, and (c) highly cited papers. The relative number of DCCPs is calculated based on Eq. 2. The three red dotted lines represent average values of $e_{i-norm}$ for papers published in each year in each group.

# DISCUSSIONS

*Scientometric discussions*

This paper demonstrates the scope of direct citations between citing publications (DCCPs) by constructing an ego-centered citation network for each paper. Different from the traditional perspective that simply counts citation relationships between a paper and its citing papers (i.e., direct citation relationships, DCRs), the current paper provides a novel method that considers citation relationships among a paper's citing papers, termed DCCPs. By utilizing a scholarly dataset from the computer science field from the Microsoft Academic Graph (MAG-CS) dataset, we find that DCCPs exist universally in medium and highly cited papers. For those papers who have DCCPs, they do occur frequently; highly cited papers tend to contain more DCCPs than others. In addition, the number of DCCPs of papers published in different years does not change dramatically.

The explorations of DCCPs have many useful applications. From the perspective of research assessment, DCCPs help scientometricians to distinguish studies that have deep/broad or dependence/independence impact [50,51]. From the perspective of citation network analysis, a certain DCCP, together with DCRs, form a "triangle" structure in the citation network. These triangles are initial steps of forming more complicated structures in the network. Various related studies can be therefore established, such as how self-citations [52,53] function in this process. From the perspective of knowledge diffusion, DCCPs assist us to understand how knowledge is diffused between "direct beneficiary." For example, in Figure 1(b), B and C can be regarded as the "direct beneficiary" of A as they both cited A. However, how knowledge/innovation is transferred between B and C remains to be unknown. Lastly, from the perspective of scholarly networks, DCCPs supplement and enrich the understanding of co-citation and bibliographic coupling relationships, as aforementioned.

Ego-centered citation networks in this paper are equivalent to citing cascades in a previous study [54]. The current study differs from our previous one (i.e., [54]) in that our previous one proposes a tool (citing cascade) to understand citation impact while the current one quantifies details of citing cascade (ego-centered citation network), such as its universality, robustness, and characteristics in different-impact papers' scenarios. In a word, our previous study is more methodology-oriented while the current one tells stories on the phenomenon itself. Meanwhile, DCCPs offer a novel aspect to examine indirect citations (co-citation, bibliographic coupling), as argued above.

*Beyond scientometrics*

DCCPs tightly relate to indirect citations, which is a typical type of indirect links more generally. The effects of indirect links have been broadly investigated outside of the field of scientometrics. For several decades, industries have been using celebrity endorsements to increase the awareness and the purchasing (or purchase intentions) of products and services. In this process, indirect links between celebrities and the public, such as links from fans to a celebrity on Facebook or Instagram, *de facto* play an important role [55], as they help more people to be aware of the products and like them, and therefore increase the probability of buying [56]. Many companies are willing to pay astronomical sums to the most well-known celebrities for endorsements. For instance, the cost of David Beckham for endorsing Adidas amounted to $160 million and Gillette $68 million [57].

Many previous studies have found a significant positive influence of indirect links on promoting growth, ranging from purchasing, reviewing, and awareness (e.g., [58,59]). Norr [60], for example, demonstrated that indirect links enabled by celebrities, defined as ties indicating that great numbers of people *know* or *follow* the celebrities, in advertising are valuable in improving various product sales in entertainment, sports, politics, food, and business. Tripp *et al.* [61] and Friedman and Friedman [62] also reported that celebrity endorsements are effective in affecting consumers to hold agreeable attitudes towards the products. The reason why these are improved is partly explained by Byrne, Whitehead, and Breen [63] who argued that a celebrity is able to build, refresh, and add new dimensions, and his/her credibility causes potential consumers to trust the products or services.

Not all studies, however, argued that indirect links always accelerate the increase of new linkages. For instance, Tom *et al.* [64] demonstrated that endorsers created by a certain company tend to create more new links (purchases) to the product than common celebrity endorsers who are not created by the company, partly because the former has been deeply embedded in the perceptions of the public with the corresponding product (company). The former is more likely to become a symbol or "spokesman" of the company. Erdogan [65] identified several potential drawbacks of celebrity endorsements, such as overshadowing the brand itself and potential for public controversy arising from events in the celebrity's life. Researchers who have contributed to this line of inquiry include Mehta [66], and Rossiter and Percy [67].

Nevertheless, indirect links are not necessarily created by celebrities (i.e., high-degree nodes in a network). Sometimes, people's decisions rely on ordinary people around them instead. Cognitive scientists, on the other hand, researched interactions between the alter-ego (e.g., direct links shown as solid lines in Figure 1(a)) and alter-alter (e.g., indirect links shown as dotted lines in Figure 1(b)), and found that the negative or positive relations between two people in a group of three are dependent on the other relations in the group [68]. Extant research has explored how such decision processes operate in other various contexts, such as power grids [69,70], bank lending systems [71], and colleague networks [72].

*Limitations and future work*

More related in-depth analysis can be conducted in the future. For example, the depth of the network, which can be defined as the length of the longest directed path from any citing paper to the owner in the network, reveals the complexity of the given network which can be used to better understand this ego-centered network, as well as for other properties, such as in- and out-degree. Moreover, the current paper analyzed DCCPs, which is one type of edge in citation networks, but did not study whether the citing papers (e.g., *B* and *D* in Figure 1(b)) are high-impact papers, which is important to understand how knowledge diffuses over time [23,73].

Yet, our conclusion cannot be generalized to any discipline outside of computer science unless future studies consider involving other domains and investigate whether the current findings also hold in fields, such as humanities and social sciences. A lack of temporal analyses constitutes another drawback of the current paper. To more elaborately support our conclusion regarding how DCCPs "help" (i.e., causal inference) or "relate to" (correlation-level analyses) accumulating citations, a temporal analysis on how a highly cited paper receives its citations over time should be implemented in the future. Particularly, the structural evolution of its ego-centered citation network can be described using indicators, such as the number of normalized DCCPs or betweenness centrality of a given citing paper (node). Achieving this could paint a more nuanced picture to understand citing behaviors and motivations [19,20].

Furthermore, the possible subject (topic) relations between papers that cite a given prior paper were not considered in the current study even though these appear to be critical. If subject connections are strong among the citing papers, then it is more likely that they will cite one another, as well. On the other hand, if the "ego" paper (a.k.a., the focal paper) is a general methodological paper in the field, e.g., a statistical test that many different topics tend to cite, or introduces a certain tool used by several empirical papers (such as VOSViewer [74]), then the citing papers will be less likely to cite one another. Perhaps if one of the citing papers becomes a highly cited paper (high visibility), then that will attract more attention and possibly more citations to the "ego" paper. Future studies could involve more pieces of information about the papers, especially the owner, in the ego-centered citation network, such as their numbers of citations (to measure papers' impact or visibility), subjects/topics, or even author-related metadata. More scientometric indicators can, therefore, be applied[2] to quantify and understand ego-centered networks.

## ACKNOWLEDGMENTS

This article is financially supported by the major program of the Social Science Foundation of China (No. 17ZDA292) and China Postdoctoral Science Foundation Funded Project (No. 2019M662729). The authors acknowledge the Indiana University Pervasive Technology Institute for providing KARST, a high-performance computing system in Indiana University, that have contributed to the research results reported within this paper. This research was supported in part by the Lilly Endowment, Inc., through its support for the Indiana University Pervasive Technology Institute, and in part by the Indiana METACyt Initiative. The Indiana METACyt Initiative at Indiana University was also supported in part by the Lilly Endowment, Inc. The authors thank Matthew Alexander Hutchinson and Xiaoran Yan for setting up empirical environments. The authors are also grateful to the anonymous reviewers for their insightful suggestions.

---

[2] For instance, the Affinity Index (AFI) and Probabilistic Affinity Index (PAI) [75,76] can be employed to understand the asymmetry and affinity of the entities (nodes) in the ego-centered network.